\documentclass[aps,pre,showpacs,nofootinbib,superscriptaddress,reprint,one,notitlepage]{revtex4}

\usepackage{amsfonts}
\usepackage{amsmath}
\usepackage{amssymb}
\usepackage{graphicx}
\usepackage[hidelinks]{hyperref}
\usepackage{color}
\usepackage{mathrsfs}
\usepackage{bbm}
\usepackage{subfigure}
\usepackage{enumerate}
\usepackage{xcolor}
\hypersetup{
	colorlinks,
	linkcolor={blue},
	citecolor={blue},
	urlcolor={blue}
}
\usepackage{times,txfonts}
\usepackage{ulem}
\usepackage{siunitx}
\usepackage{bbold}
\usepackage{sidecap}
\usepackage{mathtools}
\newcommand{\ignore}[1]{}
\newcommand{\nobibentry}[1]{{\let\nocite\ignore\bibentry{#1}}}

\newcommand{\ket}[1]{\left\vert#1\right\rangle}
\newcommand{\bra}[1]{\left\langle#1\right\vert}

\newcommand*{\rom}[1]{\expandafter\@slowromancap\romannumeral #1@}
\makeatother

\begin{document}
	
	\title{Testing a Quantum Heat Pump with a Two-Level Spin}
	
	\author{Luis A. Correa}
	\affiliation{Unitat de F\'isica Te\`orica: Informaci\'o i Fen\`omens Qu\`antics, Departament de F\'isica, Universitat Aut\`onoma de Barcelona, 08193 Bellaterra (Barcelona), Spain}
	\email{Luis.Correa@nottingham.ac.uk}
	\affiliation{School of Mathematical Sciences, The University of Nottingham, NG9 2RD Nottingham, UK}
	\author{Mohammad Mehboudi}
	\affiliation{Unitat de F\'isica Te\`orica: Informaci\'o i Fen\`omens Qu\`antics, Departament de F\'isica, Universitat Aut\`onoma de Barcelona, 08193 Bellaterra (Barcelona), Spain}

\begin{abstract}
	Once in its non-equilibrium steady state, a nanoscale system coupled to several heat baths may be thought-of as a `quantum heat pump'. Depending on the direction of its stationary heat flows it may function as e.g. a refrigerator or a heat transformer. These continuous heat devices can be arbitrarily complex multipartite systems, and yet their working principle is always the same: They are made up of several elementary three-level stages operating in parallel. As a result, it is possible to devise external `black-box' testing strategies to learn about their functionality and performance \textit{regardless} of any internal details. In particular, one such heat pump can be tested by coupling a two-level spin to one of its `contact transitions'. The steady state of this external probe contains information about the presence of heat leaks and internal dissipation in the device, and also, about the direction of its steady-state heat currents. Provided that the \textit{irreversibility} of the heat pump is low, one can further estimate its coefficient of performance. These techniques may find applications in the emerging field of quantum thermal engineering, as they facilitate the diagnosis and design optimization of complex thermodynamic cycles.
\end{abstract}


\maketitle


\section{Introduction}\label{sec:intro}
 
By `quantum heat pump', we generically mean \textit{any} stationary multi-level system simultaneously coupled to several energy sources, and capable of realizing some energy-conversion cycle, like a heat transformer or a refrigerator \cite{PhysRevLett.108.070604,PhysRevE.90.062124}. Although the interest in quantum heat pumps has skyrocketed only very recently \cite{PhysRevE.76.031105,PhysRevE.79.041129,e15062100,1310.0683v1,gelbwaser2015thermodynamics,vinjanampathy2015quantum}, these nanoscale thermodynamic devices made their \textit{debout} in scientific literature as early as 1959, when it was first suggested that a three-level maser can realize an elementary Carnot engine (or refrigerator) \cite{PhysRevLett.2.262,5123190,PhysRev.156.343}. 

The study of quantum heat pumps has proved useful in glancing at some of the central problems of quantum thermodynamics, such as understanding the potential role of `quantum effects' in nanoscale energy conversion cycles \cite{PhysRevLett.113.150402,PhysRevX.5.031044,PhysRevE.92.042126,friedenberger2015quantum}. 
Most importantly, by studying specific models of quantum heat pumps one can learn about the emergence of the laws of thermodynamics from quantum theory \cite{alicki1979engine,kosloff1984quantum,e15062100,PhysRevLett.109.090601,PhysRevE.85.061126}.

With the advent of quantum technologies, nanoscale heat pumps are finally becoming feasible, which enables many potential applications in fields as diverse as quantum information processing \cite{boykin2002algorithmic,fernandez2004algorithmic,liuzzo2015thermodynamics}, or high precision sensing \cite{PhysRevLett.110.157601,PhysRevLett.114.220405}. Several experimental realizations of quantum refrigerators and heat engines have been recently put forward \cite{PhysRevLett.108.120602,0295-5075_97_4_40003,PhysRevLett.110.256801,PhysRevLett.109.203006,PhysRevLett.114.183602,PhysRevLett.112.150602}, and even demonstrated in the lab \cite{PhysRevA.91.023431,rossnagel2015single}. In this new scenario, practical issues such as performance optimization of quantum thermodynamic cycles \cite{esposito2010universalityCA,Correa2013,1303.6558v1,PhysRevE.90.062124,PhysRevLett.112.130601,cite-key}, or the development of engineering strategies to either harnesses their \textit{quantumness} \cite{1303.6558v1,PhysRevLett.112.030602,Correa2014,gelbwaser2014heat}, or suppress undesired sources of \textit{irreversibility} \cite{PhysRevE.73.025107,PhysRevE.92.032136}, are receiving an ever-increasing attention.

In this paper we address the problem of \textit{black-box testing} of a continuous multi-level heat device. In particular, we will consider the following setting: Being supplied an unknown absorption heat pump \cite{PhysRevE.64.056130}, our aim is to learn as much as possible about its operation by interrogating an external probe. In our case, this will be a single two-level spin coupled to any of the `contact transitions' or `frequency filters' of the device \cite{PhysRev.156.343}. Our specific interests are: (i) assessing the direction of its steady-state heat flows, (ii) certifying the presence of heat leaks and internal dissipation, and (iii) estimating its `coefficient of performance', which simply gauges the cost-efficiency of its operation.

We find that the external spin deviates significantly from thermal equilibrium only when it enters in resonance with one of the open decay channels of the device. This allows to reconstruct the whole `dissipative spectrum' of the heat pump, and to break it down into its elementary constituent `stages' \cite{PhysRevE.89.042128}. For the external probe, looking effectively hotter than its surroundings is a signature of heat release from the device. On the contrary, picking up an effectively colder temperature corresponds to an incoming energy flow. Repeating this test in all three system-bath interfaces allows to tell e.g. a quantum refrigerator from a heat transformer. Additionally, the presence of \textit{detuned} elementary stages in the breakdown of the system is a clear indicator of both heat leaks and internal dissipation \cite{PhysRevE.92.032136}, and the magnitude of that detuning may be thought-of as a rough measure of the overall irreversibility. Finally, whenever the heat pump is well approximated by an \textit{endoreversible} model \cite{PhysRevE.90.062124}, one can give an estimate of its coefficient of performance based on the frequencies of the open decay channels. This black-box testing techniques may find application not only in the design optimization and diagnosis of engineered quantum thermodynamic cycles, but also in the study of the complex molecules involved in many energy-conversion biological processes.

This paper is structured as follows: In Section~\ref{sec:endo_irr} we provide a general introduction to endoreversibility, internal dissipation and heat leaks in continuous quantum thermodynamic cycles. In Section~\ref{sec:master_eq}, we present the two models of interest: On the one hand, the endoreversible three-level maser, and on the other, the irreversible four-level heat pump. We further provide the corresponding master equations accounting for the reduced dynamics of these systems when coupled to the auxiliary two-level probe. In Section~\ref{sec:testing}, we report the results of scanning the dissipative spectrum of these devices with the external two-level spin, in different regimes of operation. Finally, in Section~\ref{sec:concl} we summarize and draw our conclusions.   

\section{Endoreversible and irreversible heat devices}\label{sec:endo_irr}

\begin{figure*}
	\centering
	\includegraphics[scale=0.33]{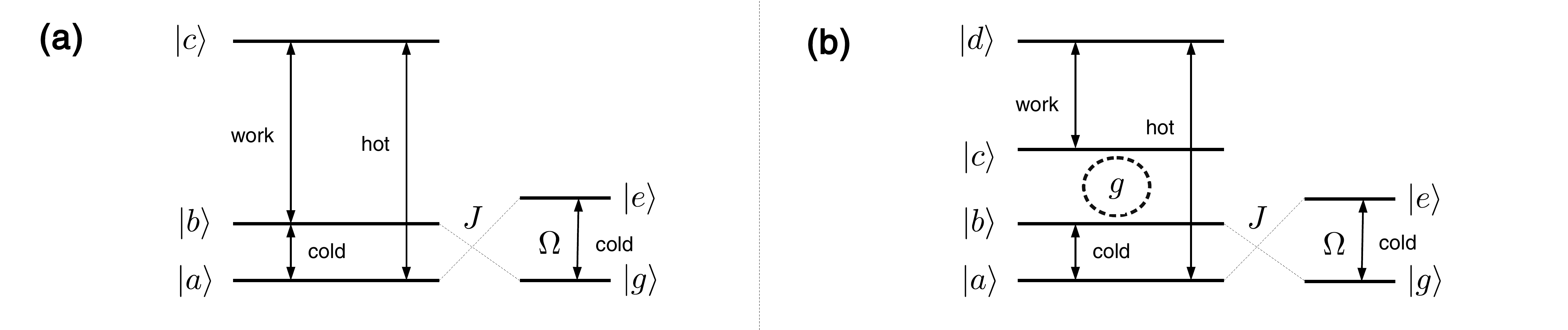}
	\caption{Schematic illustration of \textbf{(a)} a three-level heat pump and \textbf{(b)} a four-level irreversible device coupled to a tunable two-level probe of frequency $ \Omega $, with coupling strength $ J $. The three external heat baths (`work', `hot' and `cold') have temperatures $ T_w > T_h > T_c $. The arrows indicate \textit{dissipative} coupling between a given heat bath and the corresponding transition. The internal parameter $ g $ in the four level device controls its `degree of irreversibility' (see text for details).}
	\label{fig:fig1}
\end{figure*}

Back in the 1950s, the \textit{endoreversible approximation} was introduced as a simple means to model the turbines used in nuclear power generation \cite{chambadal1949thermodynamique,Yvon1955reactor,novikov1957efficiency}. Essentially, it is assumed that most of the irreversible entropy production in these systems follows from the mismatch between the temperature of the internal working fluid and the external heat sources. All other causes of irreversibility (such as heat leaks or friction) are completely ignored. Although useful in back-of-the-envelope calculations regarding performance optimization \cite{curzon1975efficiency}, the endoreversible approximation is far from satisfactory from the engineering point of view \cite{gordon1991generalized}. Indeed, friction, heat leaks, and internal dissipation do play a major role in real-life heat devices \cite{Gordon2000}. Interestingly, heat-driven three-level maser refrigerators \cite{PhysRevE.64.056130} stand as one of the few examples of a strictly endoreversible heat devices [cf. Figure~\ref{fig:fig1}(a)].

An absorption quantum heat pump or `quantum tricycle' \cite{andresen1976tricycle} operates between a hot ($ h $) and a cold ($ c $) heat bath, plus an additional heat source/repository ($ w $). In the most general case, its Hamiltonian would read
\begin{equation}
\hat H = \hat H_\text{sys} + \sum_{\alpha\in\{w,h,c\}}\hat{H}_\text{filter}^{(\alpha)} + \hat H_\text{sys-filters} + \sum_{\alpha\in\{w,h,c\}}\hat{H}_\text{filter-bath}^{(\alpha)} + \sum_{\alpha\in\{w,h,c\}} \hat H_\text{bath}^{(\alpha)} .
\label{eq:generic_heat_pump}
\end{equation}

In all what follows, we will work in units of $ \hbar=k_B=1 $. The first three terms in Equation~\eqref{eq:generic_heat_pump} stand for the multi-level system embodying the cycle. This specifically includes three two-level contact ports, or frequency filters ($ \hat{H}_\text{filter}^{(\alpha)} = \omega_\alpha\ket{1_\alpha}\bra{1_\alpha}$, where $\ket{1_\alpha} $ denotes the excited state of filter $ \alpha $). Each of these, couples to one of the unstructured \textit{bosonic} heat baths through the usual dissipative interaction term $ \hat{H}_\text{filter-bath}^{(\alpha)} = \ket{0_\alpha}\bra{1_\alpha}\otimes\hat{\mathcal{B}}_\alpha + \text{h.c.}$ \cite{breuer2002theory}. The bath operators are, in turn, $ \hat{\mathcal{B}}_\alpha = \sum_{\mu} g_{\alpha,\mu}\big(\hat a_{\alpha,\mu} + \hat a_{\alpha,\mu}^\dagger\big) $, where $ g_{\alpha,\mu}\equiv(\gamma_\alpha\omega_\mu)^{1/2} $ and $ \hat a_{\alpha,\mu} $ stand for the coupling of the mode $ \mu $ of bath $ \alpha $ to the filter, and for the corresponding annihilation operator, respectively. Finally, the last term of Equation~\eqref{eq:generic_heat_pump} corresponds to the free Hamiltonian of the heat baths.

{ It is important to remark that Equation~\eqref{eq:generic_heat_pump} is just generic template devised to encompass a broad class of systems: Actual instances of quantum heat pumps may have less terms}. For instance, in the case of the three-level maser of Figure~\ref{fig:fig1}(a), the first three terms of Equation~\eqref{eq:generic_heat_pump} reduce to { only} $ \sum_{\alpha}\hat H_\text{filter}^{(\alpha)} = \omega_c\ket{b}\bra{b} + \omega_h\ket{c}\bra{c}$ {(i.e. $ \hat{H}_\text{sys} = \hat{H}_\text{sys-filters} = 0 $)}, while $ \sum_{\alpha}\hat{H}_\text{filter-bath}^{(\alpha)} = \ket{a}\bra{b}\otimes\hat{\mathcal{B}}_c + \ket{b}\bra{c}\otimes\hat{\mathcal{B}}_w + \ket{a}\bra{c}\otimes\hat{\mathcal{B}}_h + \text{h.c.}$, that is, the frequency filters are already transitions between the energy eigenstates of the multilevel quantum heat pump. This is so also for any of the multi-stage models described in Reference~\cite{PhysRevE.89.042128}.

Working under the canonical assumptions of weak memoryless system-bath interactions, and always within the quantum optical regime, one can consistently make the Born, Markov and `rotating-wave' approximations leading to a quantum master equation in the standard Lindblad form for the reduced state $ \hat{\varrho} $ of the multi-level device (in the interaction picture with respect to $ \sum_{\alpha}\hat{H}_\text{filter-bath}^{(\alpha)} $) \cite{breuer2002theory}. That is
\begin{multline}
\frac{d\hat{\varrho}}{dt} = \sum_{\alpha\in\{w,h,c\}}\mathcal{D}_\alpha(\hat{\varrho})
\equiv\sum_{\alpha\in\{w,h,c\}}\sum_{\{\omega_\alpha\}}\Gamma_{\alpha,\omega_\alpha}\left(\hat A_{\alpha,\omega_\alpha}\hat{\varrho}\hat A_{\alpha,\omega_\alpha}^\dagger-\frac{1}{2}\{\hat A_{\alpha,\omega_\alpha}^\dagger\hat A_{\alpha,\omega_\alpha},\hat{\varrho} \}_+\right) \\
+ e^{-\omega_\alpha/T_\alpha}\,\Gamma_{\alpha,\omega_\alpha}\left(\hat A_{\alpha,\omega_\alpha}^\dagger\hat{\varrho}\hat A_{\alpha,\omega_\alpha}-\frac{1}{2}\{\hat A_{\alpha,\omega_\alpha}\hat A_{\alpha,\omega_\alpha}^\dagger,\hat{\varrho} \}_+\right).
\label{eq:general_master_eq}
\end{multline}

Equation~\eqref{eq:general_master_eq} is comprised of three \textit{dissipators} $ \mathcal{D}_\alpha $: one per heat bath. Each of them is made up of contributions coming from all the \textit{open decay channels}, at frequencies $ \{\omega_\alpha\} $. In what follows, these frequencies will be nicknamed `dissipative spectrum'. They simply label the set of transitions between energy eigenstates of the multi-level device which are effectively coupled to the heat baths. Such dissipative processes are addressed by the Lindblad or jump operators $ \hat A_{\alpha,\omega_\alpha} $ and occur at rates $ \Gamma_{\alpha,\omega_\alpha} $. Both the dissipative spectrum and the corresponding jump operators follow from the decomposition of each of the filter-bath interaction terms into a complete set of eigenoperators of the system Hamiltonian \cite{breuer2002theory}.

Note, for instance, that the three-level maser of Figure~\ref{fig:fig1}(a) features only one open decay channel per heat bath \cite{PhysRevE.64.056130}. Even if various transitions between {non-degenerate} energy eigenstates happened to be effectively coupled to a given bath, these could be jointly accounted for by a \textit{single} Lindblad operator $ \hat{A}_{\alpha,\omega_\alpha} $ and hence, only one open decay channel, provided that they all had same frequency $ \omega_\alpha $. This is the case for the family of multi-stage cycles discussed in Reference~\cite{PhysRevE.89.042128}, which are, in this sense, analogous to a (single-stage) three-level maser. On the contrary, the double-stage model depicted on the left-hand-side of Figure~\ref{fig:fig1}(b) has two \textit{distinct} open decay channels in its interfaces with both the work and the cold bath. The full master equation can be found in Reference~\cite{PhysRevE.92.032136}. {It is important to keep in mind the dissipative systems that we shall be considering are \textit{ergodic}, in the sense that no dark states exist \cite{PhysRevE.92.042123}. In particular, this ensures uniqueness of the stationary state and its independence from the initial conditions.}

The non-equilibrium steady state of the quantum heat pump ($ \hat{\varrho}_\infty $) is the stationary solution to Equation~\eqref{eq:general_master_eq}. From it, one may define the corresponding steady-state heat currents as $ \dot{\mathcal{Q}}_\alpha \equiv \text{tr}\{ \hat{H}_\text{hp}\mathcal{D}_\alpha(\hat{\varrho}_\infty) \} $, where $ \hat{H}_\text{hp} {= \hat{H}_\text{sys} + \sum_{\alpha}\hat{H}_\text{filter}^{(\alpha)} + \hat{H}_\text{sys-filters}} $ stands for the Hamiltonian of the whole multi-level heat pump \cite{alicki1979engine,kosloff1984quantum}. The thermodynamic consistency of this definition is ensured by the stationarity of the sytem's energy $ \sum_{\alpha}\dot{\mathcal{Q}}_\alpha = 0 $ (first law), and the identity $ \sum_{\alpha} \dot{\mathcal{Q}}_\alpha/T_\alpha \leq 0 $ (second law). { Let us emphasize that the non-equilibrium steady state $ \hat{\varrho}_\infty $ is maintained \textit{passively} by the heat currents and that no external driving is applied. Our irreversible entropy production rate is therefore entirely `adiabatic'} \cite{Horowitz2014}. 

It is instructive to assign an internal `spin temperature' $ \tau_{\alpha} $ \cite{abragam1958temperature} to the transitions of each decay channel, based on the steady-state populations of the corresponding ground and excited levels. For instance, in the case of the three-level maser of Figure~\ref{fig:fig1}(a), one may assign the temperature $ \tau_c = -\hbar\,\omega_{c}/[k_B\log{\big(\frac{p_b}{p_a}\big)]} $ to the transition $ \ket{a}\leftrightarrow\ket{b} $, where $ p_a $ and $ p_b $ denote the stationary populations of states $ \ket{a} $ and $ \ket{b} $. It is easy to see that there will be an incoming heat flux ($ \dot{\mathcal{Q}}_c > 0 $) whenever $ \tau_c < T_c $. On the contrary, heat will flow from the heat pump into the cold bath if $ \tau_c > T_c $. Indeed, fixing a frequency $ \omega_h $ for the transition $ \ket{a}\leftrightarrow\ket{c} $ and given a set of temperatures $ \{ T_\alpha \} $, the choice $ \omega_c < \omega_{c,\text{rev}}\equiv \omega_h T_c(T_w-T_h)/[T_h(T_w-T_c)] $ guarantees `cooling action' $ \{ \dot{\mathcal{Q}}_w > 0,\dot{\mathcal{Q}}_h < 0,\dot{\mathcal{Q}}_c > 0 \}$, whereas $ \omega_c > \omega_{c,\text{rev}} $ corresponds to the complementary `heat transformer' operation mode $ \{ \dot{\mathcal{Q}}_w < 0,\dot{\mathcal{Q}}_h > 0,\dot{\mathcal{Q}}_c < 0 \}$. 

Furthermore, there exists a configuration ($\omega_c = \omega_{c,\text{rev}}$) of the three-level maser for which \textit{all three} transitions reach local equilibrium with their corresponding heat baths (i.e. $ \tau_\alpha = T_\alpha $). As a result, the steady-state heat flows vanish, which entails an equally vanishing irreversible entropy production rate $ \dot{S} \equiv \sum_{\alpha}\dot{\mathcal{Q}}_\alpha/T_\alpha = 0 $. Once again, this is also the case for all the multi-stage models of Reference~\cite{PhysRevE.89.042128} and, in general, for any quantum heat pump with a single open decay channel per heat bath. We shall refer to all these as \textit{endoreversible} models, since their only source of irreversibility is the mismatch between `internal' and `external' temperatures \cite{PhysRevE.90.062124}. On the other hand, models with various open decay channels per {heat bath}, such as the four-level heat pump of Figure~\ref{fig:fig1}(b), are made up of \textit{detuned} elementary stages. This gives rise to two distinct irreversible processes: internal dissipation and heat leaks  \cite{PhysRevE.92.032136}, which keep energy transformation from being reversible. We refer to this type of system as \textit{irreversible} heat pumps.

From all the above, it seems clear that at the knowledge of the frequencies in the dissipative spectrum of a quantum heat pump is essential to understand its operation and the potential sources of irreversibility that may hinder it. As already advanced, in what follows we will concentrate in identifying these frequencies by means of externally probing the system-bath interfaces with a tunable two-level spin.  

\section{The quantum master equation}\label{sec:master_eq}

\subsection{The three-level maser}\label{sec:3+2}

In the next section, we will start by scanning the dissipative spectrum of the most elementary heat device: the three-level maser \cite{PhysRevLett.2.262}. Below, we detail the corresponding Lindblad master equation, including the external two-level probe. 

Following the notation of the sketch of Figure~\ref{fig:fig1}(a), the total Hamiltonian would read
\begin{multline}
\hat H_\text{total} = \sum_{\alpha\in\{w,h,c\}}\hat{H}_\text{bath}^{(\alpha)} + \left[ \omega_c\ket{b}\bra{b}+\omega_h\ket{c}\bra{c} + \Omega\ket{e}\bra{e} + J\left(\ket{a,e}\bra{b,g} + \ket{b,g}\bra{a,e}\right) \right] \\
+ (\ket{a}\bra{b} + \ket{g}\bra{e} + \text{h.c.})\otimes\hat{\mathcal{B}}_c + (\ket{b}\bra{c}+\ket{c}\bra{b})\otimes\hat{\mathcal{B}}_w 
+ (\ket{a}\bra{c}+\ket{c}\bra{a})\otimes\hat{\mathcal{B}}_h .
\label{eq:hamiltonian_3+2}
\end{multline}

Note that the two-level spin is placed in this example in the cold bath and, hence, is \textit{dissipatively} coupled to it. Notice also the `flip-flop' interaction which is established between the probe and the cold frequency filter of the device, i.e. in this case, the transition $ \ket{a}\leftrightarrow\ket{b} $. { The reference point for the notation from now on is Figure~\ref{fig:fig1}. In particular, do not be confused with the notations $ \ket{0_\alpha} $ and $ \ket{1_\alpha} $ introduced in Equation~\eqref{eq:generic_heat_pump} to account for the states of the contact ports of any generic quantum heat pump. These will not used again in what follows.} 

The eigenvalues of the maser-spin Hamiltonian, i.e. the term in brackets in Equation~\eqref{eq:hamiltonian_3+2}, are $ E_1=0 $, $ E_2 = \frac12\big(\omega_c+\Omega - \sqrt{4J^2+(\omega_c-\Omega)^2}\big)$, $E_3 = \frac12\big(\omega_c+\Omega + \sqrt{4J^2+(\omega_c-\Omega)^2}\big)$, $ E_4 = \omega_c+\Omega $, $E_5 = \omega_h $, and $E_6 = \omega_h+\Omega $, while the corresponding eigenvectors are $ \ket{1} = \ket{a,g} $, $ \ket{2} = \mathcal{N}_2^{-1}\big[(E_2-\omega_c)\ket{a,e} + J \ket{b,g}\big] $, $ \ket{3} = \mathcal{N}_3^{-1}\big[(E_3-\omega_c)\ket{a,e}+J\ket{b,g}\big] $, $ \ket{4} = \ket{b,e} $, $ \ket{5} = \ket{c,g} $, and $ \ket{6} = \ket{c,e} $ ($ \mathcal{N}_2 $ and $ \mathcal{N}_3 $ are just normalization constants).

In order to identify the necessary elements to fill in the template of Equation~\eqref{eq:general_master_eq}, we just need to decompose the terms in parenthesis from the second line of Equation~\eqref{eq:hamiltonian_3+2} as \textit{eigenoperators} of the system-filters Hamiltonian \cite{breuer2002theory}. It thus follows that the {heat source} interacts with the maser-spin system at the Bohr frequencies $ \omega_h - E_2 $, $ \omega_h - E_3 $ and $ \omega_h-\omega_c $, which correspond to jump operators $ \hat A_{w,\omega_h-E_2} = \left\langle 2 \vert b,g \right\rangle \ket{2}\bra{5} $, $\hat A_{w,\omega_h-E_3} = \left\langle 3 \vert b, g \right\rangle \ket{3}\bra{5} $, and $\hat A_{w,\omega_h-\omega_c} = \ket{6}\bra{4} $. Likewise, the hot bath contributes with three decay channels to the master equation: $ \hat A_{h,\omega_h} = \ket{1}\bra{5} $, $ \hat A_{h,\omega_h+\Omega-E_2} = \left\langle 2 \vert a,e \right\rangle \ket{2}\bra{6} $, and $ \hat A_{h,\omega_h+\Omega-E_3} = \left\langle 3 \vert a,e \right\rangle \ket{3}\bra{6} $. There are also five transitions between energy eigenstates coupled to the cold bath, namely $ \ket{1}\leftrightarrow\ket{2} $, $ \ket{3}\leftrightarrow\ket{4} $, $ \ket{1}\leftrightarrow\ket{3} $, $ \ket{2}\leftrightarrow\ket{4} $, and $ \ket{5}\leftrightarrow\ket{6} $. However, processes $ \ket{1}\leftrightarrow\ket{2} $ and $ \ket{3}\leftrightarrow\ket{4} $ have the \textit{same} associated Bohr frequency $ E_2 $. As a result, both are accounted for by a single jump operator $ \hat A_{c,E_2} = (\left\langle a,e \vert 2 \right\rangle + \left\langle b,g \vert 2 \right\rangle) \ket{1}\bra{2} + (\left\langle 3 \vert a,e \right\rangle + \left\langle 3 \vert b,g \right\rangle) \ket{3}\bra{4}$. In the case of transitions $ \ket{1}\leftrightarrow\ket{3} $ and $ \ket{2}\leftrightarrow\ket{4} $, these enter the master equation through the operator $ \hat A_{c,E_3} = (\left\langle a,e \vert 3 \right\rangle + \left\langle b,g \vert 3 \right\rangle) \ket{1}\bra{3} + (\left\langle 2 \vert a,e \right\rangle + \left\langle 2 \vert b,g \right\rangle) \ket{2}\bra{4}$. Finally, the process $ \ket{5}\leftrightarrow\ket{6} $ is represented simply by $ \hat A_{c,\Omega} = \ket{5}\bra{6} $.

The only elements of Equation~\eqref{eq:general_master_eq} which remain to be specified are the decay rates $ \Gamma_{\alpha,\omega_\alpha} $. We use the expression for the electromagnetic field in thermal equilibrium, quantized in a three-dimensional box: $ \Gamma_{\alpha,\omega_\alpha} = \gamma \omega_\alpha^3 [1+n_\alpha(\omega_\alpha)] $, where $ n_\alpha(\omega_\alpha) = (e^{\omega_\alpha/T_\alpha} - 1)^{-1}$ stands for the bosonic thermal occupation number \cite{breuer2002theory}, and $ \gamma $ denotes { the} dissipation time scale. 

Before moving to the irreversible four-level device of Figure~\ref{fig:fig1}(a), it is worth commenting on the validity of Equation~\eqref{eq:general_master_eq}. In the first place, consistency with the Born-Markov approximation demands that the system bath interaction be \textit{sufficiently} weak. More concretely, the dissipation time scale must be much larger than that of the thermal fluctuations (i.e. $ \gamma^{-1} \gg T_\alpha^{-1} $). This explicitly prevents us from working with arbitrarily cold heat sources for any given relaxation time. Similarly, due to the underlying rotating-wave approximation, there must be a large separation between the typical times of the system's evolution and the dissipation time scale (i.e $ \gamma^{-1}\gg\vert\omega_\alpha-\omega_\alpha '\vert^{-1} $). Since it is in our interest to bring the two-level spin into resonance with the open decay channels of each bath, this condition sets a lower bound for the spin-device coupling strength $ J $. Indeed, in the example above, when $ \Omega\rightarrow\omega_c $ the decay channels collapse to $ \{\omega_\alpha,\omega_\alpha\pm J \} $, which translates into $ \gamma \ll J $.

\subsection{The irreversible four-level device}\label{sec:4+2}

Besides testing an endoreversible device, we will also probe the dissipative spectrum of the simplest irreversible quantum heat pump: a detuned double stage cycle. When coupling the auxiliary spin to the cold frequency filter of the device, the total Hamiltonian reads [see Figure~\ref{fig:fig1}(b)]
\begin{multline}
\hat H_\text{total} =\omega_c(\ket{b}\bra{b}+\ket{c}\bra{c})+\omega_h\ket{d}\bra{d} + g (\ket{b}\bra{c}+\ket{c}\bra{b}) + \Omega\ket{e}\bra{e} + J\left(\ket{a,e}\bra{b,g} + \ket{b,g}\bra{a,e}\right) \\
+ (\ket{a}\bra{b} + \ket{g}\bra{e} + \text{h.c.})\otimes\hat{\mathcal{B}}_c + (\ket{c}\bra{d}+\text{h.c.})\otimes\hat{\mathcal{B}}_w 
+ (\ket{a}\bra{d}+\text{h.c.})\otimes\hat{\mathcal{B}}_h + \sum_{\alpha\in\{w,h,c\}}\hat{H}_\text{bath}^{(\alpha)}.
\label{eq:hamiltonian_4+2}
\end{multline}

The eight eigen-energies of the pump-plus-spin system [i.e. the first line in Equation~\eqref{eq:hamiltonian_4+2}] are $ E_1 = 0 $, $ E_2 = \omega_c+\Omega-g $, $ E_3 = \omega_c+\Omega+g $, $ E_4 = \omega_h $, $ E_5 = \omega_h+\Omega $, and the three real solutions of 
\begin{equation}
E_{i}^3-(\Omega+2\omega_c)E_{i}^2+(\omega_c^2+2\Omega\omega_c-g^2-J^2)E_{i}+(g^2\Omega+J^2\omega_c-\Omega\omega_c^2) = 0, \qquad i\in\{6,7,8\}.
\label{eq:polynomial}
\end{equation}

The corresponding eigenvectors write as $ \ket{1} = \ket{a,g} $, $ \ket{2} = \frac{1}{\sqrt{2}}(\ket{b,e}-\ket{c,e}) $, $\ket{3} = \frac{1}{\sqrt{2}}(\ket{b,e}+\ket{c,e}) $, $ \ket{4} = \ket{d,g} $, $ \ket{5} = \ket{d,e} $, and $ \ket{i} = \mathcal{N}_i^{-1}[((E_i-\omega_c)^2-g^2)\ket{a,e}+J(E_i-\omega_c)\ket{b,g}+gJ\ket{c,g}]$, where $ i\in\{6,7,8\} $, and $ \mathcal{N}_i $ are the corresponding normalization constants.

In this case, the work, hot and cold baths couple to five, four and ten energy transitions respectively. In general, each of these corresponds to a distinct open decay channel, since they all have different Bohr frequencies\footnote{However, at $ \Omega = \omega_c $ one finds that $ E_7 = \omega_c $ and hence, the hot decay channels at frequencies $ \omega_h $ and $ \omega_h + \Omega - E_7 $ merge.}. The Lindblad operators are ($ i \in \{6,7,8\}$)
\begin{equation*}
\hat A_{w,\omega_h-\omega_c+g}=\ket{2}\bra{5}/\sqrt{2},\,\hat{A}_{w,\omega_h-\omega_c-g}=\ket{3}\bra{5}/\sqrt{2},\,\hat{A}_{w,E_i-\omega_h}=\left\langle c,g \vert i \right\rangle\ket{4}\bra{i}
\label{eq:lindblad_4+2a}
\end{equation*}
\begin{equation*}
\hat A_{h,\omega_h}=\ket{1}\bra{4},\,\hat{A}_{h,E_i-\Omega-\omega_h}=\left\langle a,e \vert i \right\rangle\ket{5}\bra{i}
\label{eq:lindblad_4+2b}
\end{equation*}
\begin{align*}
\hat A_{c,E_i}=(\left\langle a,e \vert i \right\rangle + \left\langle b,g \vert i \right\rangle)\ket{1}\bra{i},\,\hat A_{c,E_i+g-\Omega-\omega_c}=-(\left\langle a,e \vert i \right\rangle + \left\langle b,g \vert i \right\rangle - \left\langle c,g \vert i \right\rangle)\ket{2}\bra{i}/\sqrt{2}, \\
\hat A_{c,E_i-g-\Omega-\omega_c}=(\left\langle a,e \vert i \right\rangle + \left\langle b,g \vert i \right\rangle + \left\langle c,g \vert i \right\rangle)\ket{3}\bra{i}/\sqrt{2},\,\hat{A}_\Omega=\ket{4}\bra{5}.
\end{align*}

\section{Testing a quantum heat pump}\label{sec:testing}

\subsection{Testing an endoreversible quantum heat pump}\label{sec:testing_endo}

\begin{figure*}
	\centering
	\includegraphics[width=0.45\textwidth]{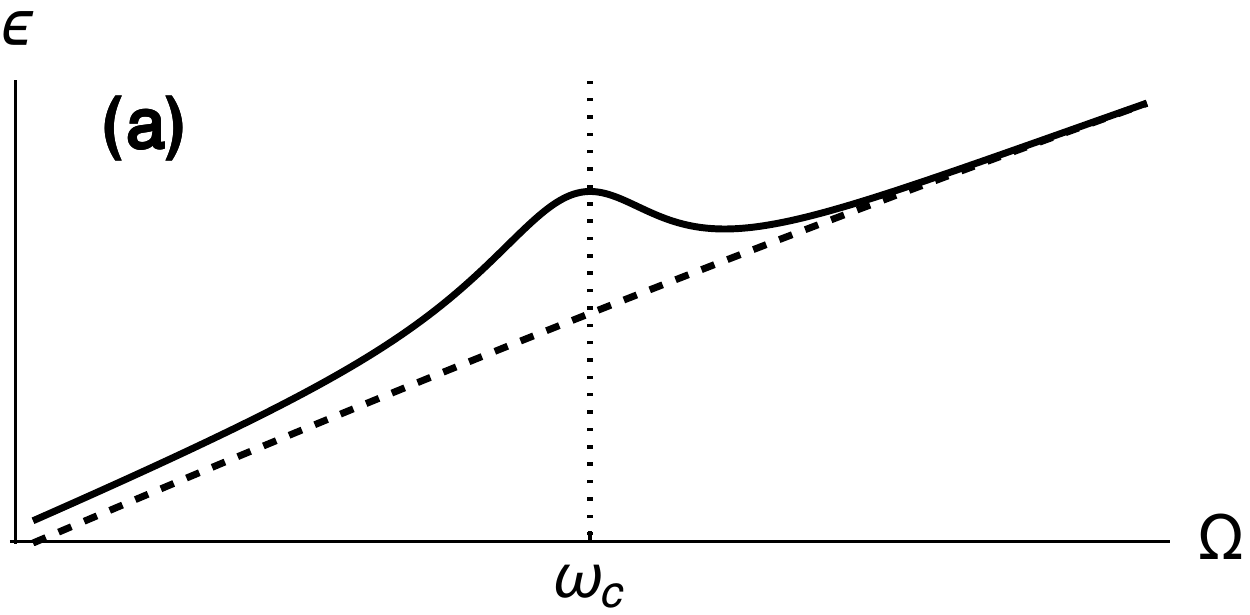}
	\includegraphics[width=0.45\textwidth]{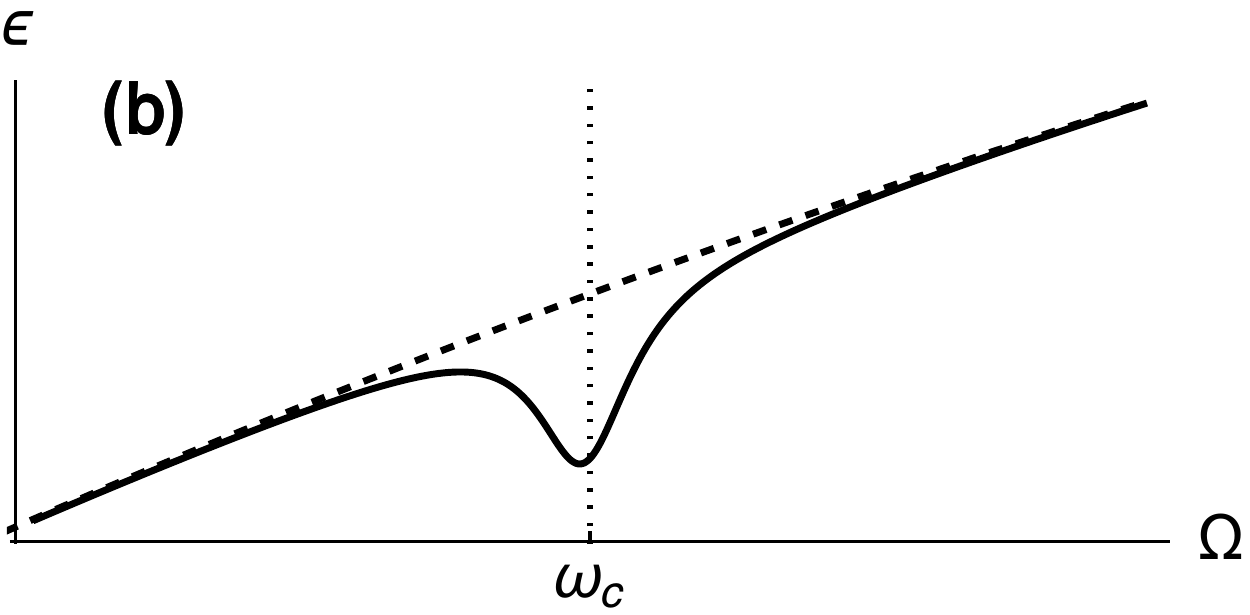}	
	\caption{(solid line) Polarization bias $ \epsilon $ of the two-level probe as a function of its frequency $ \Omega $, while scanning the `cold transition' of a three-level maser [cf. Figure~\ref{fig:fig1}(a)] \textbf{(a)} in the chiller mode ($ \omega_c = 3/4\,\omega_{c,\text{rev}} $), and \textbf{(b)} in the heat transformer mode ($ \omega_c = 5/4\,\omega_{c,\text{rev}} $). The dashed lines indicate the corresponding equilibrium polarization bias at temperature $ T_c $, and the vertical dotted lines highlight the resonance condition $ \Omega=\omega_c $. Specifically, in these plots $ T_w = 30 $, $ T_h = 20 $, $ T_c = 10 $, $ \omega_h = 40$, and $ J = 0.1 $, in units of $ \hbar = k_B = 1 $. See text for discussion.}
	\label{fig:fig2}
	\end{figure*}

We will now demonstrate how it is possible to identify the open decay channels of a multi-level quantum heat pump by externally probing it with a two-level spin. We shall start by considering the simple case of an endoreversible device, thus having a single open decay channel per heat bath.

Gathering all the pieces from Section~\ref{sec:master_eq}, it is easy to build the corresponding master equation \eqref{eq:general_master_eq}, and solve it for its non-equilibrium steady state ($ d\hat \varrho_\infty/dt = 0 $). The three-level maser can be then traced out so as to arrive at the reduced state {$ \hat{\rho}_\text{S} $} of the two-level spin. In particular, we look at the deviation of its observed polarization bias $ \epsilon $ (i.e. the difference $ \epsilon \equiv \bra{g}{\hat\rho_\text{S}}\ket{g} - \bra{e}{\hat\rho_\text{S}}\ket{e} $ between ground and excited state populations) from the equilibrium value $ \epsilon_\text{eq} = (e^{\Omega/T_c}-1)/(e^{\Omega/T_c}+1) $, as the frequency of the probe $ \Omega $ is varied. This is illustrated in Figure~\ref{fig:fig2}.

In this example, the energy exchange between the two-level spin and the heat device only becomes efficient when the probe enters in resonance with the open decay channel at $ \omega_c $. Otherwise, its dissipative interaction with the cold bath will essentially \textit{thermalize} it at temperature $ T_c $. This follows from the sharp feature in the polarization bias at $ \Omega \simeq \omega_c $ \footnote{The position of the peaks in Figure~\ref{fig:fig2} is actually slightly displaced from $ \Omega=\omega_c $ by roughly the coupling constant $ J $}, which allows to pinpoint the cold decay channel.

Furthermore, we know that the effective temperature $ \tau_c $ of the cold frequency filter of an endoreversible heat device is lower than the temperature $ T_c $ of the corresponding bath if it operates as a refrigerator, i.e. for $ \omega_c < \omega_{c,\text{rev}} $. When tuned close to resonance with the cold decay channel, the two-level spin would be affected, on the one hand, by the heat bath, and on the other, by an effectively colder two-level system at $ \tau_c < T_c$. Consequently, it will try to find a compromise between the two extremes, converging towards a steady state at some \textit{intermediate} spin temperature. The observed polarization bias of the two-level probe will be thus larger than the corresponding equilibrium value, as we can see in Figure~\ref{fig:fig2}(a). On the contrary, when the maser is working as a heat transformer, the non-equilibrium steady state of the external spin will look effectively hotter than the cold bath, and it will therefore feature a locally \textit{reduced} polarization bias, as shown in Figure~\ref{fig:fig2}(b). Hence, the polarization bias of the probe carries information not only about the frequencies of the open decay channels of the device, but also about the direction of its steady-state heat currents. As already advanced, applying this technique on all three system-bath interfaces would completely characterize the mode of operation of the heat pump.   

\subsection{Testing an irreversible quantum heat pump}\label{sec:testing_irr}

\begin{figure*}[h!]
	\centering
	\includegraphics[width=0.45\textwidth]{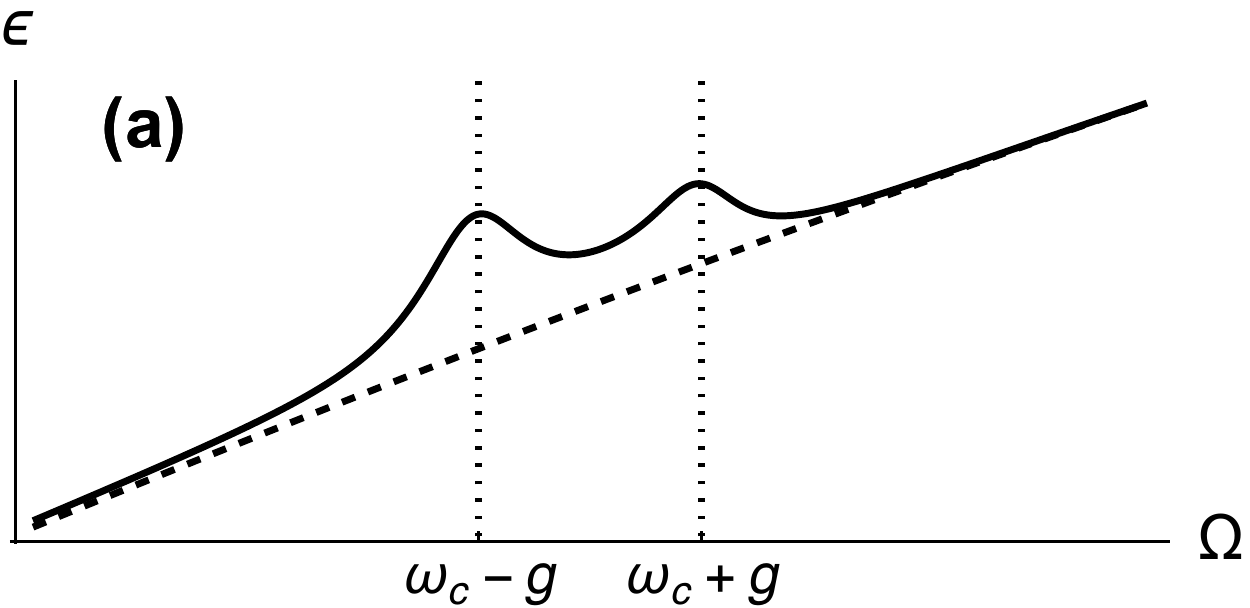}
	\includegraphics[width=0.45\textwidth]{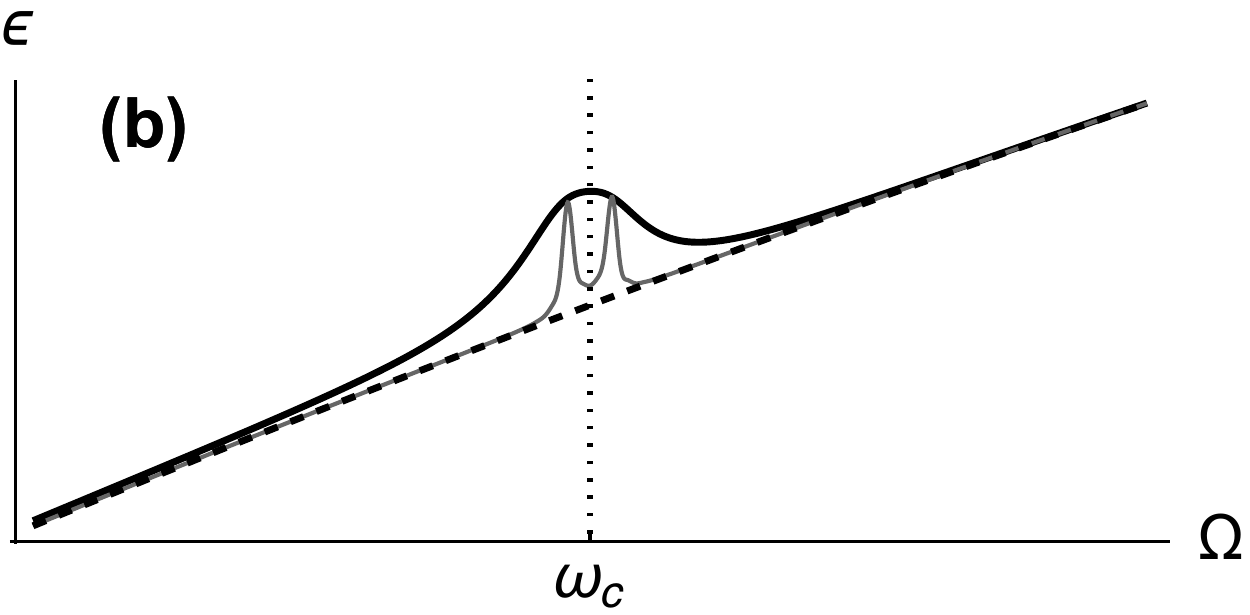}	
	\includegraphics[width=0.47\textwidth]{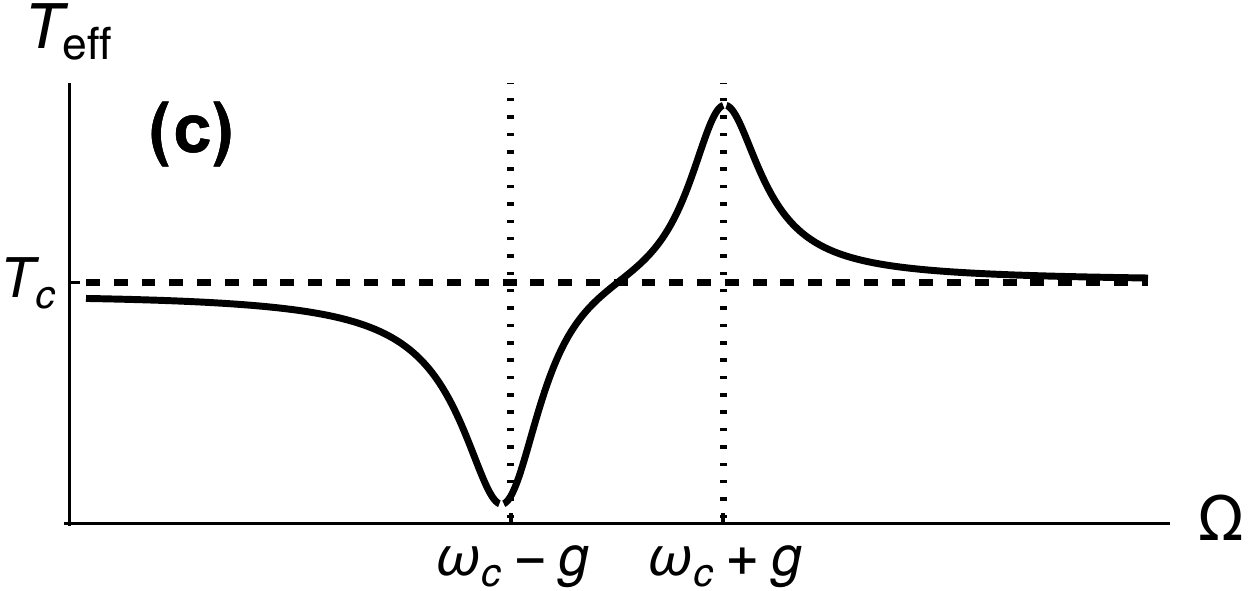}	
	\includegraphics[width=0.46\textwidth]{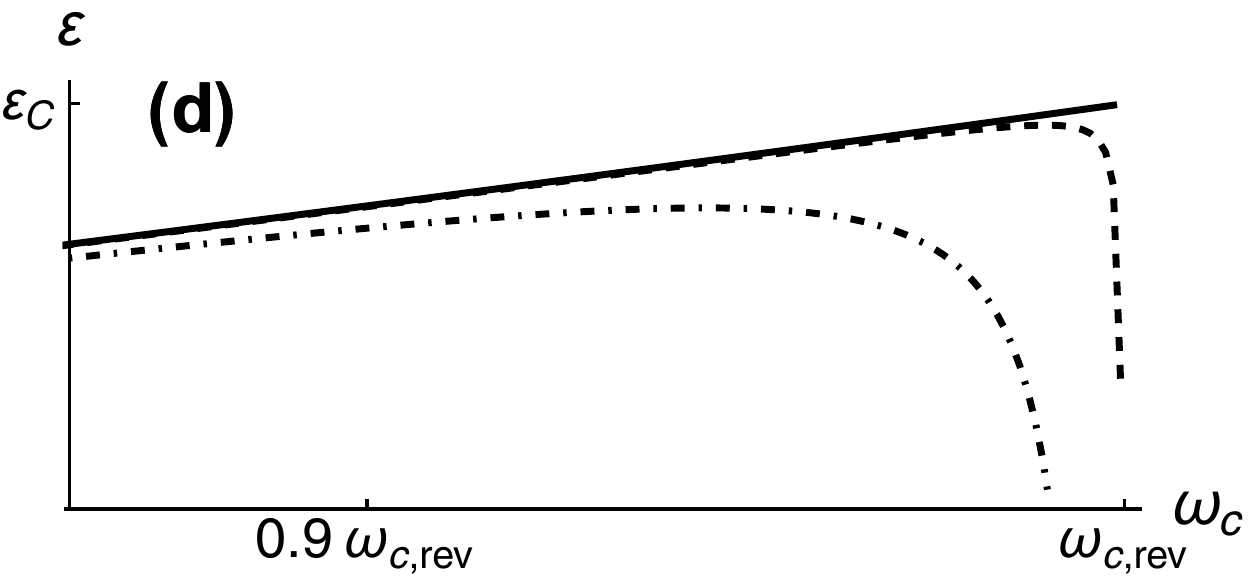}	
	\caption{\textbf{(a)} (solid black) Polarization bias $ \epsilon $ of the auxiliary two-level spin as a function of $ \Omega $, when coupled to the cold transition of an irreversible four-level chiller with $ g=0.5 $ and $ J=0.1 $. The dashed and dotted lines represent the equilibrium bias and the resonance conditions $ \Omega = \omega_c \pm g $. \textbf{(b)} Same as (a), with $ g = 0.1 $ and $ J = 0.1 $. The thin solid gray line stands for the polarization bias resulting from the weaker coupling $ J = 0.01 $. \textbf{(c)} Spin temperature of the probe $ T_\text{eff} = -\Omega/\log{[(\epsilon-1/2)/(\epsilon+1/2)]} $ versus $ \Omega $ during the scan of an irreversible four-level chiller with $ g=0.5 $ at $ \omega_c=\omega_{c,\text{rev}} $. \textbf{(d)} (solid) Coefficient of performance (COP) $ \varepsilon =\omega_c/(\omega_h-\omega_c)$ of an endoreversible chiller as a function of $ \omega_c $. The dashed and dot-dashed curves correspond to the COP of the irreversible four-level chiller, with $ g = 0.1 $ and $ g=0.5 $, respectively. All other parameters are the same as in Figure~\ref{fig:fig2}(a).
	}
	\label{fig:fig3}
\end{figure*}

Practical quantum heat pumps are likely to deviate from \textit{endoreversibility} and hence, give rise to a much richer picture. To illustrate this we will probe the cold frequency filter of an irreversible four-level chiller with our external two-level spin [see Figure~\ref{fig:fig2}(b)]. 

In general, upon probing the dissipative spectrum of any irreversible device, one expects to detect several localized features indicating the resonance with each of the existing open decay channels. Indeed, in the example of Figure~\ref{fig:fig3}(a) we can clearly locate the two decay channels at $ \omega_c \pm g $. As pointed out in the previous section, the fact that the polarization bias peaks around these frequencies indicates cooperative heat absorption from the cold heat bath by the two detuned elementary cycles conforming the heat device.

However, it may be the case that the separation between open decay channels is smaller or of the order of the spin-device coupling constant $ J $. We may then be unable to resolve the dissipative spectrum. This is illustrated by the solid black curve in Figure~\ref{fig:fig3}(b). Reducing the coupling $ J $ thus allows to increase the resolution of the scan\footnote{Recall that fixing the dissipation rate $ \gamma $ sets a lower bound on $ J $ for consistency with Equation~\eqref{eq:general_master_eq}.} and certify unambiguously the irreversibility of the heat pump. See the thin gray curve of Figure~\ref{fig:fig3}(b).

As already mentioned in Section~\ref{sec:endo_irr}, the existence of decay channels with different frequencies at the interfaces with two or more heat baths gives rise to two distinct effects: The detuning between the various elementary stages that make up the device facilitates the direct heat transfer between baths, i.e. \textit{heat leaks}, and it also results in different stationary rates of energy exchange for each of those constituent stages. As a result, for certain choices of parameters (usually close to $ \omega_c = \omega_{c,\text{rev}} $) some stages within the device may be operating as absoprtion chillers, while some others may be releasing heat into the cold bath as heat transformers. We refer to this competition as \textit{internal dissipation} \cite{PhysRevE.92.032136}. Probing an irreversible heat device may give direct evidence of internal dissipation, as it is the case in Figure~\ref{fig:fig3}(c). There, we plot the stationary spin temperature of the external probe as a function of the probing frequency $ \Omega $, precisely at $ \omega_c = \omega_{c,\text{rev}} $. We can see how the stage at frequency $ \omega_c - g $ absorbs heat from the cold reservoir, ultimately lowering the spin temperature of the resonant probe, whereas the stage at $ \omega_c + g $ is effectively hotter than its environment and thus, releases heat on average. The only reason for choosing $ T_\text{eff} $ instead of $ \epsilon $ is to graphically accentuate this effect.

\subsection{Estimating the coefficient of performance of a quantum heat pump}

So far we have been able to identify the frequencies of the dissipative spectrum of a multi-level heat device and to assess the direction of its steady-state heat currents. Probing a quantum heat pump with a two level spin also allows to give an estimate of its \textit{coefficient of performance} (COP), a figure of merit central to thermal engineering. The COP captures the cost-efficiency of the thermodynamic cycle being implemented: For instance, in the case of a quantum absorption refrigerator, it would be defined as $ \varepsilon\equiv\dot{\mathcal{Q}}_c/\dot{\mathcal{Q}}_w $, that is, the `useful effect' divided by the corresponding `energy cost'.

In an endoreversible quantum heat device, the stationary rate $ \mathcal{I} $ at which energy is exchanged with all three heat baths is identical \cite{PhysRev.156.343} and the corresponding steady-state heat currents read $ \dot{\mathcal{Q}}_\alpha = \omega_\alpha \mathcal{I} $. Consequently, the COP of e.g. an endoreversible quantum absorption chiller would be given by just the ratio of the frequencies of the cold and work contact ports $ \varepsilon = \omega_c/\omega_w = \omega_c/(\omega_h-\omega_c) $. Hence, estimating the COP of any endoreversible device only takes to probe it through two of its frequency filters so as to find the corresponding decay channels.

If the device does not deviate significantly from endoreversibility, i.e. if the frequencies of its open decay channels do not spread out too much, as it is the case in Figure~\ref{fig:fig3}(b), it will be described reasonably well by an endoreversible model. In particular, its COP may be well approximated as $ \varepsilon\simeq\omega_c/(\omega_h-\omega_c) $. This is illustrated in Figure~\ref{fig:fig3}(d), where the COP of an endoreversible three-level chiller (solid line) is compared with the actual $ \varepsilon $ of two irreversible four-level refrigerators with $ g=0.1 $ (dashed line) and $ g=0.5 $ (dot-dashed line), as calculated from the exact steady-state heat currents $ \dot{\mathcal{Q}}_\alpha $. As we can see, for the low-irreversibility setting of Figure~\ref{fig:fig3}(b), the disagreement only becomes important when operating close to the Carnot COP $ \varepsilon_C = \frac{T_c(T_w-T_h)}{T_w(T_h-T_c)}$, while for the more irreversible case portrayed in Figure~\ref{fig:fig3}(a), the approximation is much worse, in general. Note as well that the slight shifts of the peaks/wells in the polarization bias, due to the finite spin-device coupling $ J $, are an additional source of error in the estimation of the COP.

\section{Conclusions}\label{sec:concl}

In this paper we have shown how black-box testing of a multi-level quantum heat device can provide relevant information such as the direction of its steady-state heat currents, its degree of irreversibility, and even a good estimate of its coefficient of performance. All this information may be obtained by simply looking at the departure from thermal equilibrium of the steady state of an external two-level probe coupled to one of the contact transitions of the heat device. Crucially, in this testing protocol, no information about the internal structure of the heat device is required.

In particular, it is possible to identify the frequencies of the open decay channels at the interface of the system and each of the heat baths, and thus, to characterize the various elementary thermodynamic cycles that make up the quantum heat pump. We emphasize that all this is possible without knowledge of the Hamiltonian of the multi-level heat device. In some cases, the competition between these elementary stages, which is a signature of internal dissipation, may be directly witnessed.

For our analysis, we have considered the dissipative dynamics of an absorption three-level maser, which realizes the simplest endoreversible thermodynamic cycle, when dissipatively coupled to our external two-level spin. We have further studied the case of a four-level absorption chiller as the paradigm of the more complex irreversible heat devices.

Our results provide a clear and intuitive picture of the inner workings of generic heat-driven thermodynamic cycles, \textit{beyond specific models}, and they may find applications in the design of cost-efficient nano-engineered thermodynamic devices, or in the characterization of the complex energy-conversion processes taking place in mesoscopic biological systems.

\section*{Acknowledgments}

We are thankful to Jos\'{e} P. Palao and Gerardo Adesso for carefully reading and commenting on this manuscript. We gratefully acknowledge financial support from Spanish MINECO (Project No. FIS2013-40627-P) and Generalitat de Catalunya (CIRIT Project No. 2014 SGR 966). L.A.C acknowledges funding from the European Research Council (ERC) Starting Grant GQCOP (Grant No. 637352). M.M. acknowledges financial support from EU Collaborative Project TherMiQ (Grant Agreement 618074).







\bibliographystyle{mdpi}
\renewcommand\bibname{References}

\end{document}